\documentclass[aip,apl,reprint,twocolumn,superscriptaddress]{revtex4-1} 

\usepackage{amsmath,amsthm,amsfonts,amssymb,verbatim,comment}
\usepackage[mathscr]{euscript}
\usepackage{graphicx}
\usepackage{url}
\usepackage[normalem]{ulem}
\usepackage[T1]{fontenc}
\usepackage[cp1250]{inputenc}

\usepackage{dcolumn}%
\usepackage{bm}%

\newcommand{\val}[2]{\ensuremath{#1~\mathrm{#2}}}

\begin{document}

\title{High data-rate atom interferometer for measuring acceleration}

\author{Hayden J. McGuinness}
\email[Corresponding author: ]{hmcgui@sandia.gov}
\affiliation{Sandia National Laboratories, Albuquerque, NM 87185, USA}
\author{Akash V. Rakholia}
\affiliation{Sandia National Laboratories, Albuquerque, NM 87185, USA}
\affiliation{Center for Quantum Information and Control (CQuIC), Department of Physics and Astronomy, University of New Mexico, Albuquerque, New Mexico, 87131, USA} 
\author{Grant W. Biedermann}
\affiliation{Sandia National Laboratories, Albuquerque, NM 87185, USA}
\affiliation{Center for Quantum Information and Control (CQuIC), Department of Physics and Astronomy, University of New Mexico, Albuquerque, New Mexico, 87131, USA}

\date{\today}

\begin{abstract}
We demonstrate a high data-rate light-pulse atom interferometer for measuring acceleration. The device is optimized to operate at rates between 50 Hz to 330 Hz with sensitivities of $0.57\; \mathrm{\mu g/\sqrt{Hz}}$ to $36.7\; \mathrm{\mu g/\sqrt{Hz}}$, respectively. Our method offers a dramatic increase in data rate and demonstrates a path to new applications in highly dynamic environments. The performance of the device can largely be attributed to the high recapture efficiency of atoms from one interferometer measurement cycle to another.
 
\end{abstract}
  
\maketitle

  
Since the inception of the light-pulse atom interferometer (LPAI) in 1991 \cite{Riehle91,Kasevich91}, the field has matured to the point where atom interferometers are poised to significantly advance applications in gravimeter surveys \cite{pyramid}, seismic studies, inertial navigation \cite{McGuirk,Canuel06,Wu,Durfee}, and tests of fundamental physics \cite{gravwave,Durr,Fixler,HMuller}. However, for atom interferometer accelerometers to compliment or replace conventional technologies in the more dynamic of these applications, the rate at which the acceleration measurements are taken, the data rate, equal to the inverse of the interferometer cycle time, must be on the order of 100 Hz or more. To date, all reported LPAI accelerometer demonstrations have operated at a rate on the order of a few Hertz or less. In this letter, we report an LPAI that can operate at data rates up to 330 Hz with sensitivities suitable for many of the aforementioned applications.

Since the interferometer response scales quadratically with temporal duration, typical approaches prefer low data rates using long interrogation times to achieve the highest sensitivity.  However, at short interrogation times it is likely that atom interferometers will still maintain high sensitivity, long-term accuracy, and the ability to perform both rotation and acceleration measurements. This has stimulated interest in short interrogation LPAI \cite{Butts}, and encourages the exploration of new configurations that trade sensitivity for data rate and reduced system demands. We investigate the scenario where the gravitational displacement during the time of flight (TOF) is less than the magneto-optical trap (MOT) size, corresponding to 10 ms for a 1 mm diameter cloud. Having a small displacement allows us to recapture the interferometer atoms with high efficiency and drastically reduce the measurement dead time associated with replenishing the trap atoms. This enables an increase in data rate by one to two orders of magnitude and achieves short-term sensitivities comparable to other demonstrations targeting field use \cite{pyramid,Canuel06}.
 
 \begin{figure}\centering{
\includegraphics[width=1\columnwidth]{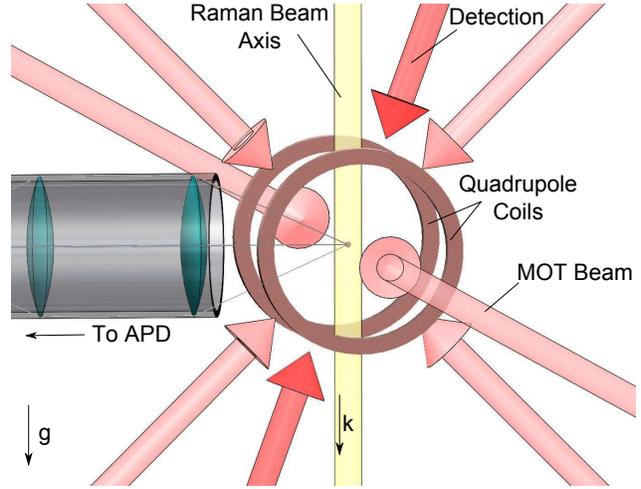}
\caption{\label{fig:MZ} Diagram of the interferometer setup. The quadrupole coils are 32 mm in diameter.} }
\end{figure}



A particularly successful class of LPAIs, and that which is studied here, are those that use stimulated Raman transitions in a $\pi/2 - \pi - \pi/2$ pulse sequence \cite{Berman}. This sequence uses stimulated photon recoil to coherently separate the atomic wavepackets spatially, redirect the wavepackets toward one another, and finally interfere the formerly separate wavepackets. In a sufficiently uniform gravitational field, the dominant phase shift comes from interaction with the light \cite{Bongs}. Each pulse imprints its spatial phase $\phi_i$ on the atoms, inducing a cumulative phase shift $\Delta \phi = \phi_1 - 2\phi_2 + \phi_3$ between the ground and excited states. The value of this phase shift determines the relative population of the two states upon detection through $P = 1/2(1-\cos(\Delta\phi))$. This phase shift is proportional to the acceleration of the atoms along the effective wavevector $\textbf{k}=\textbf{k}_\textbf{1} - \textbf{k}_\textbf{2}$ such that $a_k=\frac{\Delta \phi}{k T^2}$, where $T$ is the free evolution interrogation time between pulses. Hence through the measurement of the ground and excited state populations, $\Delta \phi$ is measured, and the acceleration $a_k$ can be determined. We use this technique to measure local gravity and our device's inherent sensitivity by propagating the Raman fields parallel to gravity.
 


%

The LPAI takes place in a miniature quartz vacuum cell having outer dimensions of \val{14 \times 16 \times 80}{mm}$^3$ with a pressure of $2 \times 10^{-9}$ Torr (see Figure \ref{fig:MZ}). Within this cell we load a traditional MOT at a rate of $4 \times 10^{7}$ atoms/s from a background vapor of Rb 87.  The axial magnetic field gradient of the quadrupole coils is calculated to be 7.8 G/cm. The trapping beams at 780 nm are red detuned from the $F=2 \rightarrow F'=3$ cycling transition by \val{9}{MHz}. The beams are collimated, having a $1/e^2$ diameter of \val{5.2}{mm} and combined peak intensity of 182 mW/cm$^2$ (saturation parameter of $s_0 = 108$) after passing through the cell.  We optimize the MOT parameters for high recapture efficiency and short recapture duration. To prevent accumulation in the dark state $F=1$, 0.5 mW of repump light resonant with the $F=1 \rightarrow F'=2$ transition is directed at the atoms.


At the start of the interferometer cycle, the MOT atoms are sub-Doppler cooled to approximately \val{5.5}{\mu K}. A \val{1.5}{G} bias field is then applied along the axis of the Raman beams to split the Zeeman degeneracy and define a quantization axis. The atoms are then prepared in the $F=1$ manifold with a depump pulse of \val{100}{\mu s} duration resonant with the $F=2 \rightarrow F'=2$ transition. This state preparation distributes the population in the $F=1$ manifold with 43 percent in the $m=0$ sublevel and the rest nearly evenly distributed in the $m=\pm 1$ sublevels. Consequently, a large bias field is used to shift the $\pm1$ sublevels out of resonance with the interferometer beam tuned to the clock transition $F=1,m=0 \rightarrow F=2,m=0$. Even though the atoms in these levels do not participate in the interferometer and add a small amount of background noise during detection, they are easily recaptured, allowing for shorter atom collection time. After depumping, the atoms are interrogated with the Raman beams.

\begin{figure} \centering{
\includegraphics[width=1\columnwidth, clip=true]{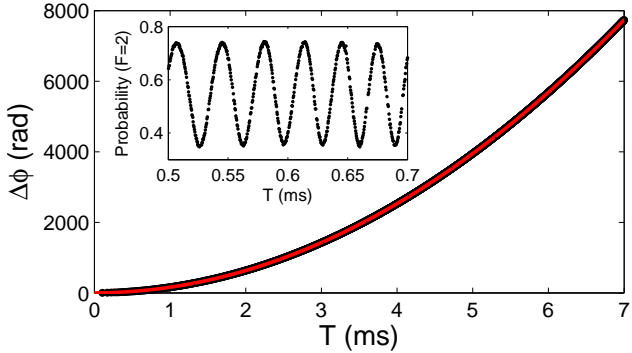}
\caption{\label{fig:gravity} Gravity value determination through the measurement of phase shift versus interrogation time. The data (black) is fit to a chirped sinusoid, of which the mid-fringe intercepts are shown, and yields a value of $g=9.7916378 \pm 9.7\times 10^{-6} \textrm{m/s}^2$. The inset shows a portion of the raw data, renormalized to exclude the signal from spectator atoms, where its chirped nature is apparent.}}
\end{figure}

The Raman laser is seeded by an external cavity diode laser locked \val{1.23}{GHz} red of the $F=2 \rightarrow F'=3$ transition. We generate two phase-coherent laser fields separated in frequency by the hyperfine splitting by injection locking two uncoated laser diodes with the output of a fiber electro-optic modulator driven at 6.835 GHz \cite{Xue}. One of the diodes is tuned to predominately amplify the positive sideband while the other is tuned to amplify the carrier. The two beams are then coupled onto orthogonal axes of the same polarization maintaining fiber, delivered to the sensor-head, then separated and directed so that the positive sideband field is incident down upon the atoms and the carrier field is incident up upon the atoms.  The beams are collimated to a $1/e^2$ diameter of \val{5.5}{mm}, and demonstrate a Rabi frequency of $\Omega_{eff}=2\pi \times 161$ kHz. The pulse duration is controlled with a field-programmable  gate array having 20 ns resolution. We ramp the frequency of the positive sideband at a rate of $\mathrm{\textbf{k} \cdot \textbf{g}} = 2\pi \times 25.1\; \mathrm{kHz/ms}$ to compensate for the Doppler shift the atoms experience due to gravitational freefall.

Following the interferometer sequence, the population in $F=2$ and the total atom number are detected by light induced fluorescence from two sequential detection pulses, each \val{100}{\mu s} in duration. During the pulses, 1.2 percent of the fluorescence is collected into an avalanche photodiode (Hamamatsu c5460-1). The first pulse consists of a detection beam resonant with the $F=2 \rightarrow F'=3$ cycling transition, thus measuring the population in $F=2$. The second pulse consists of the detection beam and the repump beam, measuring the total population, including the spectator atoms. The detection beam is retro-reflected to balance the scattering force. The atoms are then recaptured in the MOT and the cycle is immediately repeated. The interferometer operates in a continuous loop, recapturing the atoms from one interferometer measurement for use in the next. Loading from background vapor replaces lost atoms and we find an average equilibrium atom number of $\approx 2 \times 10^{5}$ after optimizing for sensitivity.

The most important factor enabling a high data-rate is efficient atom recapture, which allows for the majority of the atoms, approximately 95 to 85 percent depending on the data rate and TOF, to be reused from cycle to cycle. With this method, the equilibrium atom number is achieved in 1 to 2 ms. The recapture duration is consistent with a calculated characteristic restoring time of $\approx 3.5$ ms for our MOT parameters \cite{Metcalf}. Attaining a similar atom number loading from only vapor would take on the order of 20 ms in our system, limiting our maximum theoretical data rate to approximately 50 Hz. Efficient recapture is made possible due to sufficient sub-Doppler cooling and the atoms only falling a few hundred microns during the relatively short TOF, so that most of them never travel outside the MOT trap volume.  We observe that higher rates with shorter TOFs lead to an even smaller displacement from the trap center, increasing recapture efficiency.

\begin{figure} \centering{
\includegraphics[width=1\columnwidth]{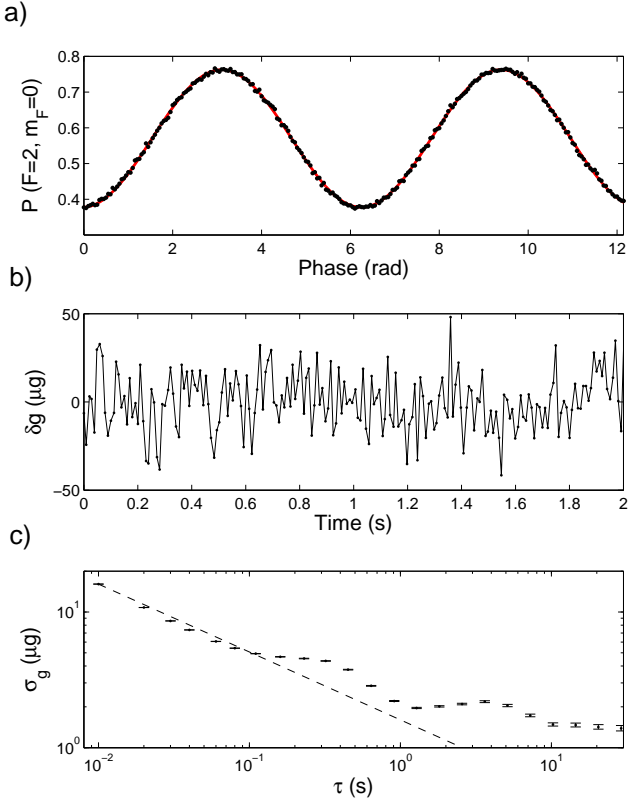}  
\caption{\label{fig:100Hz} 100 Hz case study. (a) Fringe pattern with no averaging, renormalized to exclude the offset from spectator atoms with interrogation time $T=3.415$ ms. The red sinusoidal fit is a guide for the eye. A $\pi$ phase shift corresponds to an acceleration of 1.71 mg.  (b) A section of the time record on the side of a fringe. The quantity $\delta$g is the change in the acceleration from mean value. (c) The two-sample Allan deviation as a function of time with error bars. The dotted line shows a $\tau^{-1/2}$ trend , starting from the 10 ms Allan deviation value.}}
\end{figure}


To verify the interferometer response to accelerations, we determine the value of gravity in the lab by measuring the interferometric phase shift as a function of the interrogation time (which was varied from 0 to 7 ms, see Figure \ref{fig:gravity}). For the purpose of this experiment, we do not compensate for the Doppler shift the atoms experience, which leads to a chirped sinusoid with linear and quadratic terms of the form P = $A(T)\mathrm{cos}(\phi_0\,+\,\tau_{\pi} kg_k(1+2/\pi)T \,+\, k g_k T^2)\, +\, C$, where $A(T)$ is a slowly varying envelope function, $\tau_{\pi}$ is the $\pi$ pulse duration, $g_k$ is gravity, and $\phi_0$ is an arbitrary phase factor \cite{Peters}. The term linear in $T$ is due to the time-variation of the Doppler shift during the Raman pulses, and is therefore proportional to the Raman pulse duration and gravity. 

The fit of our data yields a value of $g_k=9.7916378 \pm 9.7\times 10^{-6} \textrm{m/s}^2$, where the uncertainty is one standard deviation from the central value. This value agrees well with the nearest surveyed value after accounting for elevation change and the uncertainty in the collinearity of \textbf{k} and gravity \cite{NGS}. This shows that in a static situation with a $T$ up to 7 ms we can predict the behavior at the level of our sensitivity without Doppler compensation. To achieve maximal signal (the Doppler shift reduces the fringe visibility), we compensate for the Doppler shift in the remainder of our measurements.



As a sample case, we examine the sensitivity of our device at 100 Hz, a data rate which represents a favorable trade-off in terms of being fast enough for many navigation applications but which also exhibits good sensitivity. Interferometer parameters such as the sub-Doppler cooling and recapture durations were optimized to achieve the greatest sensitivity. Figure \ref{fig:100Hz}(a) shows a fringe pattern obtained by scanning the relative phase of the Raman beams between pulses. To attain good statistics, 12,000 data points were taken mid-fringe. We use the 2-sample Allan deviation of the acceleration signal to determine the shot-to-shot sensitivity $\sigma_g= \frac{\delta \phi_g }{ g_k k}\frac{1}{\mathrm{T^2}}$, where $\delta \phi_g$ is the uncertainty in the value of $\Delta \phi$ shot-to-shot (see Figures \ref{fig:100Hz}(a) and \ref{fig:100Hz}(b)). We report the short term sensitivity accounting for data rate by $\sigma_s= \sigma_g/ \sqrt{R}$, where $R$ is the data rate, which yields $1.6 \;\mathrm{\mu g/\sqrt{Hz}}$ in this case. As expected, the Allan deviation generally decreases with increasing time, although there are two noticeable deviations from this trend centered at about 0.3 and 4 seconds. The technical noise, discussed below, of the system causes these departures and will likely not limit long term stability in future devices.

\begin{figure} \centering{
\includegraphics[width=1\columnwidth, clip=true]{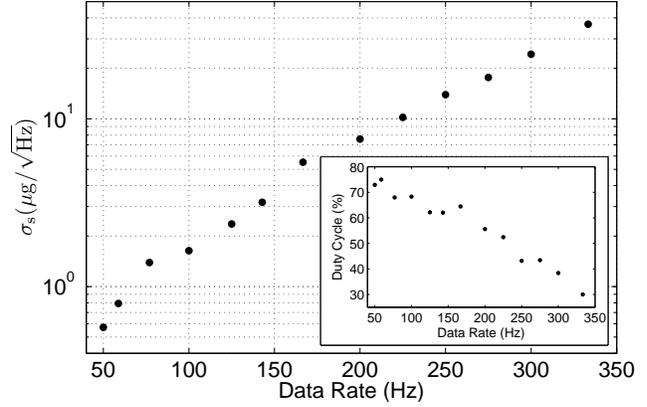}
\caption{\label{fig:SenVSbw} Plot of short-term sensitivity versus data rate, the inverse of the interferometer cycle time, from 50 Hz to 330 Hz. The inset plots the duty cycle versus data rate. We optimize the duty cycle for sensitivity.} }
\end{figure}

Using the same methods as in the 100 Hz case, we explore the performance of our device for data rates of 50 Hz to 330 Hz. Figure \ref{fig:SenVSbw} shows the achieved sensitivity, while the inset plots the duty cycle (the ratio of the total interrogation time, 2$T$, and cycle time), for that range. The sensitivity monotonically increases with increasing date rate, having values of $0.57 \mathrm{\mu g/\sqrt{Hz}}$ to $36.7 \mathrm{\mu g/\sqrt{Hz}}$ at 50 Hz and 330 Hz, respectively. The duty cycle varies from 75 to 30 percent from 50 Hz to 330 Hz, respectively. At high data rates the optimal sub-Doppler and recapture durations make up a larger fraction of the cycle time, leaving less time for interrogation.  Furthermore, the increase in data rate only partially compensates for the reduced interrogation time causing a reduction in sensitivity.

The quantity $\delta \phi_s$ as a function of data rate is nearly constant, with an average value of $31.1 \mathrm{mrad/shot}$, and dropping only approximately 25 percent from 50 Hz to 330 Hz. This suggests the sources of noise were relatively insensitive to data rate. Two sources in particular account for the bulk of the noise: Raman beam phase instability and magnetic field noise. By beating the two Raman beams together on a photodiode, the Raman beam phase noise was estimated to be $21\; \mathrm{mrad/shot}$, which we largely attribute to optical component instability. By analyzing the background magnetic field with a magnetometer, the noise due to the ambient magnetic field was estimated to be $15\; \mathrm{mrad/shot}$, most likely caused by the second-order Zeeman effect. Adding these in quadrature leads to a combined noise of $26\; \mathrm{mrad/shot}$ for these sources, the bulk of the measured noise.


In conclusion, we have demonstrated a high data-rate atom interferometer operating at rates between 50 Hz and 330 Hz having sensitivities suitable for many gravimeter, seismic, and inertial navigation applications. The key to achieving this performance at high data rate and small sensor-head size comes from optimizing the system for high atom recapture efficiency. In particular, a recapture efficiency of between 95 and 85 percent allowed for a significantly shorter loading time than in many other LPAI systems. Nevertheless, a theoretical analysis suggests that our achieved sensitivity is roughly an order of magnitude worse than the ultimate noise floor limit given by quantum projection noise. Readily implemented improvements such as enclosing the sensor head in a magnetic shield to block ambient magnetic fields and increasing the stability of the Raman beam optical setup could significantly reduce the overall noise and allow our LPAI to achieve a sensitivity at the level of $100\; \mathrm{n g/\sqrt{Hz}}$ at 100 Hz.

\begin{acknowledgments}
We gratefully thank G. Burns, K. Fortier, T. Loyd, and Y.-Y. Jau for their contributions. This work was supported by the Laboratory Directed Research and Development (LDRD) program at Sandia National Laboratories.
\end{acknowledgments}


\providecommand{\noopsort}[1]{}\providecommand{\singleletter}[1]{#1}%

\end{document}